# Large-Area Two-Dimensional Layered MoTe$_2$ by Physical Vapor Deposition and Solid-Phase Crystallization in a Tellurium-Free Atmosphere

**Jyun-Hong Huang[1], Kuang-Ying Deng[2], Pang-Shiuan Liu[1], Chien-Ting Wu[3], Cheng-Tung Chou[2], Wen-Hao Chang[4,5], Yao-Jen Lee[3,6], and Tuo-Hung Hou[1,5 a]**

[1] Department of Electronics Engineering and Institute of Electronics, National Chiao Tung University, Hsinchu 300, Taiwan
[2] Department of Chemical and Materials Engineering, National Central University, Jhongli 320, Taiwan
[3] National Nano Device Laboratories, Hsinchu 300, Taiwan
[4] Department of Electrophysics, National Chiao Tung University, Hsinchu 300, Taiwan
[5] Taiwan Consortium of Emergent Crystalline Materials, (TCECM), Ministry of Science and Technology, Taipei 106, Taiwan
[6] Department of Physics, National Chung Hsing University, Taichung 402, Taiwan

**Abstract**

Molybdenum ditelluride (MoTe$_2$) has attracted considerable interest for nanoelectronic, optoelectronic, spintronic, and valleytronic applications because of its modest band gap, high field-effect mobility, large spin-orbit-coupling splitting, and tunable 1T′/2H phases. However, synthesizing large-area, high-quality MoTe$_2$ remains challenging. The complicated design of gas-phase reactant transport and reaction for chemical vapor deposition or tellurization is nontrivial because of the weak bonding energy between Mo and Te. Here, we report a new method for depositing MoTe$_2$ that entails using physical vapor deposition followed by a post-annealing process in a Te-free atmosphere. Both Mo and Te were physically deposited onto the substrate by sputtering a MoTe$_2$ target. A composite SiO$_2$ capping layer was designed to prevent Te sublimation during the post-annealing process. The post-annealing process facilitated 1T′-to-2H phase transition and solid-phase crystallization, leading to the formation of high-crystallinity few-layer 2H-MoTe$_2$ with a field-effect mobility of ~10 cm$^2$/(V·s), the highest among all nonexfoliated 2H-MoTe$_2$ currently reported. Furthermore, 2H-MoS$_2$ and Td-WTe$_2$ can be deposited using similar methods. Requiring no transfer or chemical reaction of metal and chalcogen reactants in the gas phase, the proposed method is potentially a general yet simple approach for depositing a wide variety of large-area, high-quality, two-dimensional layered structures.

Correspondence and requests for materials should be addressed to T-H Hou

E-mail: thhou@mail.nctu.edu.tw

1. **INTRODUCTION**

Among the numerous actively researched two-dimensional (2D) layered transition-metal dichalcogenides (TMDs),[1-5] MoTe$_2$ is one of the most intriguing but less explored materials.[6-8] MoTe$_2$ is the only TMD that can be directly synthesized in both semiconducting 2H and semimetallic 1T′ phases[8-11] because of the small energy difference between the two phases (~40 meV[8,12]). Bulk 2H-MoTe$_2$ films have an indirect band gap of 0.93 eV, whereas monolayer 2H-MoTe$_2$ has a direct band gap of 1.1 eV.[7] A band gap close to that of Si is highly desirable for numerous electronic and optoelectronic applications such as phototransistors,[13] field-effect transistors (FETs),[6,14-16] and tunnel FETs.[17] Furthermore, outstanding carrier transport properties have been reported for MoTe$_2$. A theoretical acoustic phonon-limited mobility of more than 2500 cm$^2$/(V·s) for 2H-MoTe$_2$[18] and an experimental Hall mobility of 4000 cm$^2$/(V·s) for 1T′-MoTe$_2$[8] are among the highest in 2D TMDs.[18] Additionally, MoTe$_2$ has a larger spin-orbit-coupling (SOC) splitting of the valence band than MoS$_2$ or MoSe$_2$ because of the heavier Te atom (238 meV for MoTe$_2$ vs 161 meV for MoS$_2$ and 175 meV for MoSe$_2$[19]). This large SOC splitting energy in addition to inversion-symmetry breaking in odd-layer TMDs results in unique valley- and spin-dependent selection rules in the material and its long-lived spin and valley polarization. This may provide a route toward the integration of valleytronics and spintronics.[6,20,21] Recently, 1T′-MoTe$_2$ has been predicted to be a Weyl semimetal candidate that exhibits salient quantum phenomena,[22-24] including extremely large magnetoresistance and pressure-driven superconductivity.[23] Furthermore, the phase transition of MoTe$_2$ from 2H to 1T′ was previously induced by applying laser irradiation[25] or lateral tensile strain[12] because of the relatively low transition energy barrier of 0.88 eV between the two phases.[12] Local phase engineering was used to form ohmic homojunctions and reduce contact resistance in MoTe$_2$ transistors.[25]

Despite the promising properties of MoTe$_2$, the development of MoTe$_2$ applications has been hindered largely by its less matured large-scale synthesis process compared with that of other widely studied TMDs such as MoS$_2$, MoSe$_2$, WS$_2$, and WSe$_2$. The difficulty of synthesizing large-scale 2D MoTe$_2$ stems from the weak bonding energy between Mo and Te,[9] which results in low chemical reactivity during the formation of MoTe$_2$. Additionally, MoTe$_2$ is relatively unstable and easily decomposes at high temperatures in a Te-deficient atmosphere.[9] Furthermore, achieving a homogeneous crystalline phase is challenging because of the small energy difference between the 2H and 1T′ phases. High-quality 2D MoTe$_2$ is most often produced by exfoliating the bulk material that is synthesized using chemical vapor transport (CVT) methods.[6-8,12-15] Despite being useful for researching fundamental material properties, the lack of necessary large-area and layer-by-layer controllability limits the potential of exfoliation methods for practical applications. Recently, several techniques have been developed for synthesizing large-area MoTe$_2$. Few-layer MoTe$_2$ was grown using molecular beam epitaxy employing Mo and Te as sources in an ultrahigh vacuum. However, the small grain size of the resulting MoTe$_2$ led to relatively poor transport properties that were dominated by defect hopping.[26] Chemical vapor deposition (CVD) through gas-phase reactions between transition metal and chalcogen reactants from solid sources has been widely adopted for preparing high-quality, large-grain sulfide and selenide monolayers,[27,28] but a stable CVD process for synthesizing MoTe$_2$ remains elusive because the weak bonding energy between Mo and Te complicates the design of their gas-phase transport and reaction. A well-designed mixed precursor compound of MoO$_3$, MoCl$_5$, and Te was reported to facilitate the CVD MoTe$_2$ synthesis,[11] but the layer controllability and large-area uniformity were unfavorable. Post-tellurization of Mo feedstock on a substrate, including Mo or MoO$_3$ deposited using electron-gun evaporation or sputtering[9,10] or ammonium heptamolybdate through drop casting[29], is currently the most reliable method for synthesizing homogeneous MoTe$_2$ on centimeter-scale substrates. Nevertheless, few of these studies synthesized high-quality MoTe$_2$ that was suitable for FET devices. The only field-effect mobility value reported (1 cm$^2$/(V·s)) was relatively low.[9] The mobility in MoTe$_2$ obtained using the CVT and exfoliation methods typically ranges from 0.3 to 40 cm$^2$/(V·s).[8,13,14] Furthermore, precisely controlling the Te partial vapor pressure is crucial during the synthesis of single-phase, high-crystallinity MoTe$_2$.[9,10] Empirical designs for the temperature gradient and gas flow in the furnace, necessary to achieve a desirable vapor pressure on the substrate, are often unreliable and difficult to reproduce. The possibility of film decomposition further complicates the design of the process and reduces the process window.

This study reports a new method for depositing large-area, highly crystalline, few-layer 2D MoTe$_2$ without the need of gas-phase transport or reaction of the Mo and Te reactant species. Both Mo and Te were directly deposited on the substrate by using physical vapor deposition (PVD) of a sputtered MoTe$_2$ target, followed by post-deposition annealing at 650 °C in a Te-free atmosphere. To prevent the MoTe$_2$ decomposition and Te sublimation at high temperatures, a SiO$_2$ capping layer was deposited prior to the post-deposition annealing. The SiO$_2$ capping layer also served as a 2D-confined encapsulation that facilitated 1T′-to-2H phase transition and solid-phase crystallization under an excessive Te supply. The high crystallinity of 2H-MoTe$_2$ was confirmed using Raman spectroscopy, X-ray photoelectron spectroscopy (XPS), and high-resolution tunneling electron microscopy (HRTEM). The fabricated 2H-MoTe$_2$ back-gated FET exhibited p-type conduction characteristics with a current on/off-ratio of $10^5$ and a field-effect mobility of ~10 cm$^2$/(V·s), the highest among all the nonexfoliated, large-area 2D layered MoTe$_2$ currently reported. We also demonstrated that the developed direct PVD method is applicable for depositing 2H-MoS$_2$ and Td-WTe$_2$ by simply replacing the sputter target, and is thus a potential general method for depositing a wide variety of large-area, high-quality 2D layered materials.

2. RESULTS

2.1 2H-MoTe$_2$ Deposition

The method used to produce large-area, high-crystallinity, multilayer 2D MoTe$_2$ is illustrated in Figure 1a. MoTe$_2$ was directly deposited on a 3 cm × 4 cm SiO$_2$ substrate by using DC magnetron sputtering and a MoTe$_2$ target. The phase of the MoTe$_2$ target was confirmed to be 2H through an exfoliation method (Supporting Information Figure S1). The as-deposited MoTe$_2$ thin film had low crystallinity according to the Raman measurement detailed in Figure S2 (Supporting Information). The films were then immediately capped with a 50-nm evaporated SiO$_2$ layer, followed by a 50-nm SiO$_2$ layer deposited using plasma-enhanced chemical vapor deposition (PECVD). The capped MoTe$_2$ samples were annealed in a low-pressure, 2-in. hot-wall furnace at 650 °C for 3–24 h in N$_2$ to improve their crystallinity. A more detailed discussion of the MoTe$_2$ deposition is provided in the Methods Section. The SiO$_2$ capping layer was crucial during the post-annealing process. Without the capping layer, the as-deposited, disordered Mo–Te composite completely decomposed at 650 °C in N$_2$. Similar decomposition was reported for a sputtered MoS$_2$ film annealed in a sulfur-deficient atmosphere.[30] The bilayer capping used was designed to prevent the outdiffusion of Te vapor during annealing. PECVD SiO$_2$ is dense and of favorable quality, but the evaporated SiO$_2$ layer prevents potential plasma damage or oxidation on the as-deposited films during PECVD. The SiO$_2$ capping layer serves as a 2D-confined encapsulation of the Mo–Te composite and facilitates the formation of 2D layered MoTe$_2$ under a sufficient Te supply during annealing.

Raman spectrum of the 2D layered MoTe$_2$ films annealed for 24 h are presented in Figure 1b. The peaks at 171.4 ($A_{1g}$), 234.7 ($E^1_{2g}$), and 290.0 ($B^1_{2g}$) cm$^{-1}$ corresponded to the major Raman vibrational modes of 2H-MoTe$_2$.[8,31] The relatively low intensity of the out-of-plane $A_{1g}$ peak was attributed to the Raman excitation wavelength at 532 nm.[7] The out-of-plane $B^1_{2g}$ peak is known to vanish in monolayer and bulk MoTe$_2$ and is characteristic for few-layer MoTe$_2$ because of the translation symmetry breaking along the c-axis.[31] The difference between the out-of-plane $A_{1g}$ and in-plane $E^1_{2g}$ peaks decreased when the thickness of the MoTe$_2$ films was increased (Supporting Information Figure S3).[31] The inset in Figure 1b displays the uniform color contrast of the MoTe$_2$ film deposited on the SiO$_2$ substrate. Raman spectra were collected for a total of 60 sites on the sample. Figure 1c presents the statistics of the major Raman peaks of these spectra. Variation in the wavenumber of the $A_{1g}$, $E^1_{2g}$ and $B^1_{2g}$ peaks was small, indicating excellent large-area uniformity. The attainable sample size in this work was limited by the diameter of the annealing furnace tube.

## 2.2 Time Evaluation of 1T′-to-2H Phase Transition

The growth mechanism of 2H-MoTe$_2$ during the post-annealing process warranted further investigation. Figure 2a–d present top-view optical microscope images of MoTe$_2$ annealed for periods ranging from 3 to 24 h. Raman spectroscopy was performed to identify the regions with 2H (Figure 2e) and 1T′ (Figure 2f) phases. In the 3-h sample, MoTe$_2$ was mostly in the 1T′ phase, identified by the Raman peaks at 163 (B$_g$) and 262 (A$_g$) cm$^{-1}$.[8] 2H-MoTe$_2$ was identified only in sparse and isolated circular islands, possibly centered around nucleation sites. The boundaries of 1T′ and 2H MoTe$_2$ phases were visible because of their color contrast under the optical microscope. The circular 2H-MoTe$_2$ regions gradually expanded outward with annealing time and merged to form a complete 2H-MoTe$_2$ layer after 24 h. A similar 1T′-to-2H phase transition was previously reported when 2H-MoTe$_2$ was synthesized through the tellurization of Mo metal.[10] Additionally, we observed visible metallic gray precipitates located mostly at the 1T′/2H-MoTe$_2$ phase boundaries (Figure 2(b-d)), which exhibited Raman peaks at 122 and 143 cm$^{-1}$ (Figure 2g), in agreement with those obtained from bulk Te flakes (Supporting Information Figure S4).[32,33] Maintaining an excessive-Te atmosphere was discovered to be essential for stabilizing stoichiometric 2H-MoTe$_2$ phases.[10] The alloy phase diagram of 2D MoTe$_2$[8] reveals that excessive Te favors the formation of 2H-MoTe$_2$ at 1T′/2H phase boundaries. Therefore, Te migration and segregation may play a crucial role in the phase transition from 1T′ to 2H during annealing.

**2.3 Physical Characterization of 2H-MoTe$_2$**

The material stability of the few-layer 2H-MoTe$_2$ was further investigated. After the post-annealing process, the 100-nm SiO$_2$ cap could be safely removed using diluted hydrogen fluoride (HF, 0.58%) for 8 min without substantially affecting the Raman spectrum of the 2H-MoTe$_2$ film (Supporting Information Figure S5). This decapping process is essential for subsequent material analysis and device fabrication. The decapped film remained stable in a vacuum desiccator for more than 2 months (Supporting Information Figure S6). Figure 3 presents XPS analysis of the decapped 2H-MoTe$_2$. The characteristic Mo–Te binding energies were present at 228.2 (Mo 3d$_{5/2}$), 231.4 (Mo 3d$_{3/2}$), 572.9 (Te 3d$_{5/2}$), and 583.3 eV (Te 3d$_{3/2}$).[9] The calculated full width at half maximum values were 0.69 eV for Mo 3d$^{5/2}$ and 0.92 eV for Te 3d$^{5/2}$. The overall composition was 30.3% of Mo and 69.7% of Te. The Te-rich composition was the average over the entire XPS beam size of several square micrometers and could be attributed to the aforementioned localized Te precipitates because the 100-nm SiO$_2$ capping prohibited Te from sublimation during annealing. A more stoichiometric composition with less Te precipitates might be achieved by further optimizing the sputtering and capping processes.

The atomic structure of MoTe$_2$ was investigated using HRTEM. The cross-sectional image in Figure 4a reveals three-layer stacking of 2H-MoS$_2$ (annealed for 24 h) with a thickness of 2.1 nm in total or 0.7 nm for each layer, which is consistent with those obtained by exfoliation[6,7], CVD[11], or post-tellurization[9] methods. The plan-view image in Figure 4b and its fast-Fourier-transform filtered image in Figure 4c illustrate a perfect hexagonal lattice wherein the Mo and Te atomic arrangement corresponds to the 2H phase. Selected-area electron diffraction (SAED) with a beam diameter of 650 nm revealed a single set of diffraction patterns with six-fold symmetry (Figure 4d). The result indicates that the grain size of the single-crystal 2H-MoTe$_2$ was at least more than 650 nm. By contrast, 1T′-MoTe$_2$ (annealed for 3 h) exhibited a rather disordered lattice arrangement (Figure 4e). The SAED pattern when a beam diameter of 260 nm was used (Figure 4f) confirms that the 1T′-MoTe$_2$ obtained using a short annealing time was polycrystalline. Therefore, the post-annealing process not only facilitated the 1T′-to-2H phase transition but also significantly improved the crystallinity of MoTe$_2$.

**2.4 Electrical Characterization of 2H-MoTe$_2$**

The decapped 2H-MoTe$_2$ was transferred to another SiO$_2$ (300 nm) substrate before using it to fabricate back-gated transistors because the SiO$_2$ on the original substrate could have been degraded during the post-annealing process as a result of the potential mixing or diffusion of Mo and Te atoms. Pd (25 nm) was deposited for use as the source/drain contacts. The effect of different contact metals was discussed in Supporting Information (Figure S7). Figure 5a illustrates the typical transfer characteristics (drain current ($I_D$)–gate voltage ($V_G$) sweep) of one of the MoTe$_2$ devices. The device channel was 8 μm long. The device exhibited p-type conduction characteristics with a current on/off-ratio of $10^5$ at a drain voltage ($V_D$) of −1 V. Strong $V_D$ dependence in the subthreshold region is not expected in conventional long-channel FETs, but rather a characteristic of Schottky-barrier FETs (SB-FETs) with a substantial barrier height.[34] In SB-FETs, the width of the reverse-bias source-side Schottky barrier and thus the hole field-emission (tunneling) current are modulated by the drain bias.[15] Similar strong $V_D$ dependence has also been reported in other MoTe$_2$ transistors with either Ti/Au[6,9,14] or Pd[15] contacts. The field-effect hole mobility $\mu_h$ in the linear region was extracted using $\mu_h = (dI_D/dV_G)(L/W)(1/V_D C_G)$ at $V_D = -1$ V, where $L$, $W$, and $C_G$ represent the channel length, channel width, and gate capacitance per unit area, respectively. The extracted $\mu_h$ of ~10 cm$^2$/(V·s) was an underestimation because of the presence of the substantial Schottky barriers at contacts. Nevertheless, the calculated $\mu_h$ was comparable to that of the mechanically exfoliated MoTe$_2$[8,13,14] and significantly higher than that obtained using post-tellurization.[9] This result indicates that long-time (24 h) post-annealing might produce superior MoTe$_2$ films (e.g., larger grains, fewer defects) compared with the short-time (1 h) post-tellurization.[9] Figure 5b displays the well-behaved output characteristics ($I_D$–$V_D$ sweep) of the p-type MoTe$_2$ device. Although the $I_D$–$V_D$ curves appeared to be linear at a small $V_D$ bias, this should not be interpreted as the absence of Schottky barriers.[34]

3. DISCUSSION

3.1 Growth Mechanism and Solid-phase Crystallization

Figure 6 presents a summary of the plausible recrystallization and phase transition mechanism of 2D MoTe$_2$ deposited using PVD followed by post-annealing. The low-crystallinity Mo–Te composite deposited using sputtering was completely encapsulated by the substrate and the SiO$_2$ capping layer to prevent Te outdiffusion or sublimation during the subsequent post-annealing process (Figure 6a). In the 2D-confined encapsulation, the post-annealing process provided sufficient energy for the lateral diffusion of atoms and facilitated the initial recrystallization of polycrystalline 1T′ MoS₂. Why the 1T′ phase growth was preferred over the 2H phase during the initial annealing is still unclear. 1T′-MoTe₂ was reported to become more thermodynamically stable under a lateral tensile strain of <3%,[12] and such a strain could have been potentially induced by the substrate and capping layer. Because the as-deposited Mo–Te composite was Te-rich, the excessive Te segregated into the remaining amorphous regions between the 1T′ grains (Figure 6b) and provided favorable Te-rich conditions for the 2H-MoTe₂ nucleation (Figure 6c).[8-10] When the annealing time was increased, the regions of recrystallized 2H-MoTe₂ expanded outward from nucleation sites along the excessive-Te forefront by transforming 1T′-MoTe₂ into 2H-MoTe₂, producing circular 2H-MoTe₂ grains (Figure 6d). Finally, the entire layer was recrystallized into the 2H phase through the merging of isolated 2H-MoTe₂ grains, and the excessive Te precipitated at the grain boundaries (Figure 6e). Beginning from a small number of nucleation sites, this recrystallization process is similar to the solid-phase crystallization process widely used for recrystallizing amorphous Si into large-grain poly-Si by using low-temperature and long-time annealing processes.[35,36] Our proposed mechanism suggests that reducing the amount of excessive Te in the as-deposited film while retaining a slight surplus of Te could reduce the number of nucleation sites and thus further enlarge the 2H-MoTe₂ grain size at long post-annealing times. It would also suppress the amount of Te precipitated at the grain boundaries and thus achieve a more stoichiometric film.

**3.2 Depositing TMDs beyond MoTe₂**

In addition to depositing high-quality MoTe₂, we successfully deposited other 2D-layered TMDs by using similar PVD methods with SiO₂ capping and chalcogen-free post-annealing, except for using different sputtering targets (see Methods Section). Figure 7a presents the Raman spectrum of 2H-MoS₂ with characteristic Raman peaks at 385.6 ($E_{2g}^1$) and 406.6 ($A_{1g}$) cm$^{-1}$.[37,38] Figure 7b displays the Raman spectrum of Td-WTe₂ with characteristic Raman peaks at 109.9 ($A_2^4$), 121.9 ($A_1^9$), 131.7 ($A_1^8$), 166.2 ($A_1^5$), and 216.1 ($A_1^2$) cm$^{-1}$.[39,40] These results suggest that the proposed method is capable of depositing arbitrary 2D layered structures other than MoTe₂.

Generally, conventional CVD methods are separately optimized for each TMD material or even different substrate size because complex processes and hardware designs are required to transfer reactant species in the gas phase to the substrate while simultaneously facilitating favorable chemical reactions. By contrast, both the transition metals and chalcogens are directly deposited onto the substrate through sputtering in the proposed method, which thus involves neither reactant transfer nor chemical reactions in the gas phase and is not restricted by numerous process constrains such as the choice of precursors, a proper ratio between the vapor pressures of metal and chalcogen reactants, the deformation of synthesized films, or turbulent flow control. Therefore, this deposition method greatly simplifies precursor, process, and hardware design and requires only facilities already used in the semiconductor industry such as sputtering systems, electron-gun evaporators, PECVD chambers, and annealing furnaces.

## 4. CONCLUSION

In summary, we reported a new method for depositing large-area, highly crystalline, few-layer 2H-MoTe$_2$ without the need of gas-phase transport or reaction of the Mo and Te reactant species. Both Mo and Te were physically deposited onto the substrate by sputtering a MoTe$_2$ target. A composite SiO$_2$ capping layer was designed to prevent Te sublimation during the post-annealing process in the Te-free atmosphere. The post-annealing process facilitated 1T′-to-2H phase transition and solid-phase crystallization, leading to the formation of large-grain 2H-MoTe$_2$ with high field-effect mobility comparable to that of the mechanically exfoliated MoTe$_2$. We also demonstrated that the developed PVD method is applicable for depositing 2H-MoS$_2$ and Td-WTe$_2$ by simply replacing the sputter target. Requiring no transfer or chemical reaction of metal and chalcogen reactants in the gas phase, the proposed method is potentially a general yet simple route for depositing a wide variety of large-area, high-quality 2D layered structures and could be used in future mass production and device applications.

## 5. EXPERIMENTAL SECTION

*MoTe₂ Deposition:* A DC magnetron sputtering system with a base pressure of $10^{-8}$ Torr was used for few-layer MoTe₂ deposition and employed a commercial MoTe₂ target (99.9% purity; Kojundo Chemical Laboratory). MoTe₂ was deposited on a 3 cm × 4 cm SiO₂ substrate (300-nm SiO₂ on Si) for 30 s at 400 °C using an Ar flow of 20 sccm, a working pressure of 10 mTorr, and DC power of 50 W. The as-deposited MoTe₂ thin films were then immediately transferred to an electron-gun evaporator and capped with a 50-nm-thick SiO₂ film by using a source of SiO₂ granules. An additional 50-nm-thick SiO₂ layer was deposited on top of the evaporated SiO₂ through PECVD using SiH₄ and N₂O as precursors at 300 °C. The capped MoTe₂ samples were then annealed in a low-pressure, 2-in. hot-wall furnace in a N₂ atmosphere. The temperature of the furnace was increased at a rate of 37.5 °C/min from 200 °C to 650 °C, and the pressure of the furnace was maintained at 30 Torr. The samples were annealed at 650 °C for 3 to 24 h, after which the temperature cooled naturally to 200 °C.

*Physical Characterization of MoTe₂:* The lattice vibrational modes were identified using Raman spectroscopy with a laser source of wavelength 532 nm (Finder Vista, Zolix Instruments Co. Ltd). The system was calibrated using the Raman peak of Si at 520 cm$^{-1}$. The laser power was maintained at 5 mW. The chemical composition was determined using XPS (Thermo Fisher Scientific Theta Probe, Al Kα X-ray source). The nanostructure of the few-layer MoTe₂ in the plan and cross-sectional views was investigated using HRTEM (plan view: JEOL TEM-2100F/EDS; cross-sectional view: FIB-FEI Nova600, TEM-FEI Tecnai G2 F20). To prepare the sample for the plan-view HRTEM, the MoTe₂ sample on the SiO₂ substrate was coated with a PMMA (950 PMMA A4, MICROCHEM) layer by using a spin coater at 3000 rpm for 60 s. After baking at 100 °C for 2 min, the sample was immersed in a dilute HF solution (1.2%) at room temperature for 15 min. Subsequently, the MoTe₂ thin film with PMMA coating was gently peeled off the SiO₂/Si substrate in water and transferred to a lacey carbon film on a mesh copper grid. Finally, the PMMA was removed using acetone and the sample was thoroughly rinsed with water.

*Fabrication of Back-Gated MoTe₂ FET:* The three-layer MoTe₂ film was transferred to a 300-nm-thick SiO₂ substrate on Si by using a transfer process similar to that used for preparing the HRTEM sample. The 300-nm SiO$_2$/Si was used as the back-gate dielectric/electrode, respectively. The source/drain regions were first defined using conventional photolithography. Pd of thickness 25 nm was deposited using electron-gun evaporation and lifted off in an acetone solution. The active channel region was then patterned using conventional photolithography and plasma dry etching of Ar + CHF$_3$. The characteristics of the back-gated FETs were measured at room temperature using an HP-4156B semiconductor parameter analyzer.

*MoS$_2$ and WTe$_2$ Deposition:* The MoS$_2$ and WTe$_2$ deposition processes were similar to that for MoTe$_2$, unless otherwise mentioned. For depositing MoS$_2$, a MoS$_2$ target (99.9 % purity; UMAT) was sputtered at 400 $^{\circ}$C for 10 s on a sapphire substrate. The SiO$_2$-capped MoS$_2$ sample was annealed at 780 $^{\circ}$C for 12 h. For depositing WTe$_2$, a WTe$_2$ target (99.9 % purity; Toshima Manufacturing Co., Ltd) was sputtered at 400 $^{\circ}$C for 30 s on a sapphire substrate. The SiO$_2$-capped WTe$_2$ sample was annealed at 750 $^{\circ}$C for 1 h.

**SUPPORTING INFORMATION**

Supporting Information is available from the Wiley Online Library or from the author.

**ACKNOWLEDGEMENTS**

This work was supported by the Ministry of Science and Technology of Taiwan under grant: MOST 103-2221-E-009-221-MY3, NSC 102-2119-M-009-002-MY3, MOST 105-2119-M-009-014-MY3, the Asian Office of Aerospace Research and Development (AOARD) under grant: 16IOA013, and the Office of Naval Research Global (ONRG) under grant: N62909-17-1-2022. T.-H. Hou acknowledges support by NCTU-UCB I-RiCE program, under grant MOST 105-2911-I-009-301. The authors are grateful to the Nano Facility Center at National Chiao Tung University and National Nano Device Laboratories, where the experiments in this paper were performed.

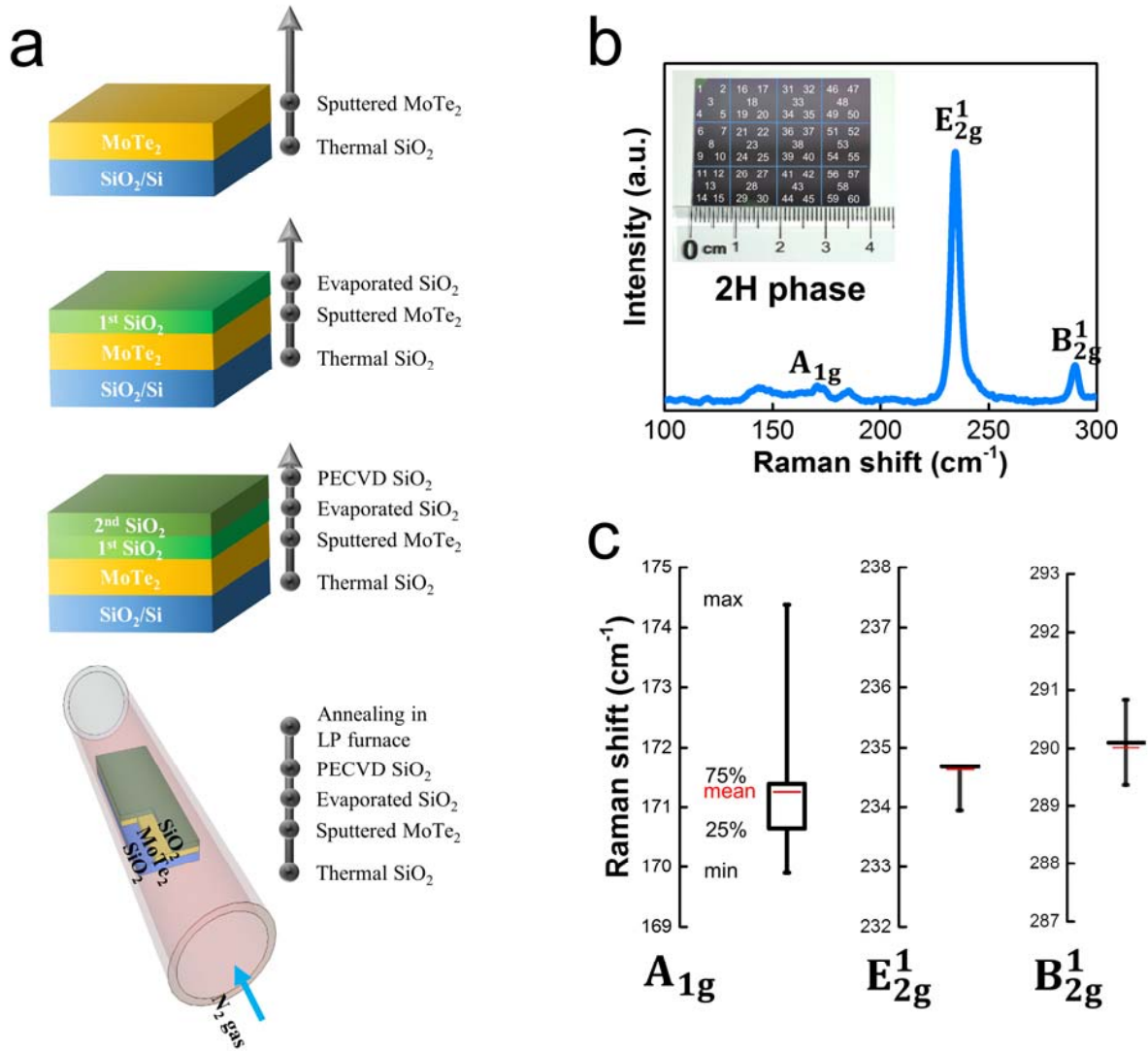

**Figure 1** | (a) Schematic of the PVD process for depositing large-area, high-quality MoTe₂. A MoTe₂ thin film was sputtered directly onto a SiO₂ substrate by using a MoTe₂ target, after which a composite evaporation−PECVD SiO₂ capping layer was deposited. The sample was then subjected to a post-annealing process at 650 °C in N₂ by using a low-pressure furnace. (b) Raman spectrum of 2H-MoTe₂ annealed for 24 h. Inset: uniform color contrast of MoTe₂ on a 3 cm × 4 cm SiO₂ substrate. (c) Statistics of major Raman peaks— $A_{1g}$, $E_{2g}^1$, and $B_{2g}^1$—determined from 60 sites on the substrate as labelled in the inset of (b).

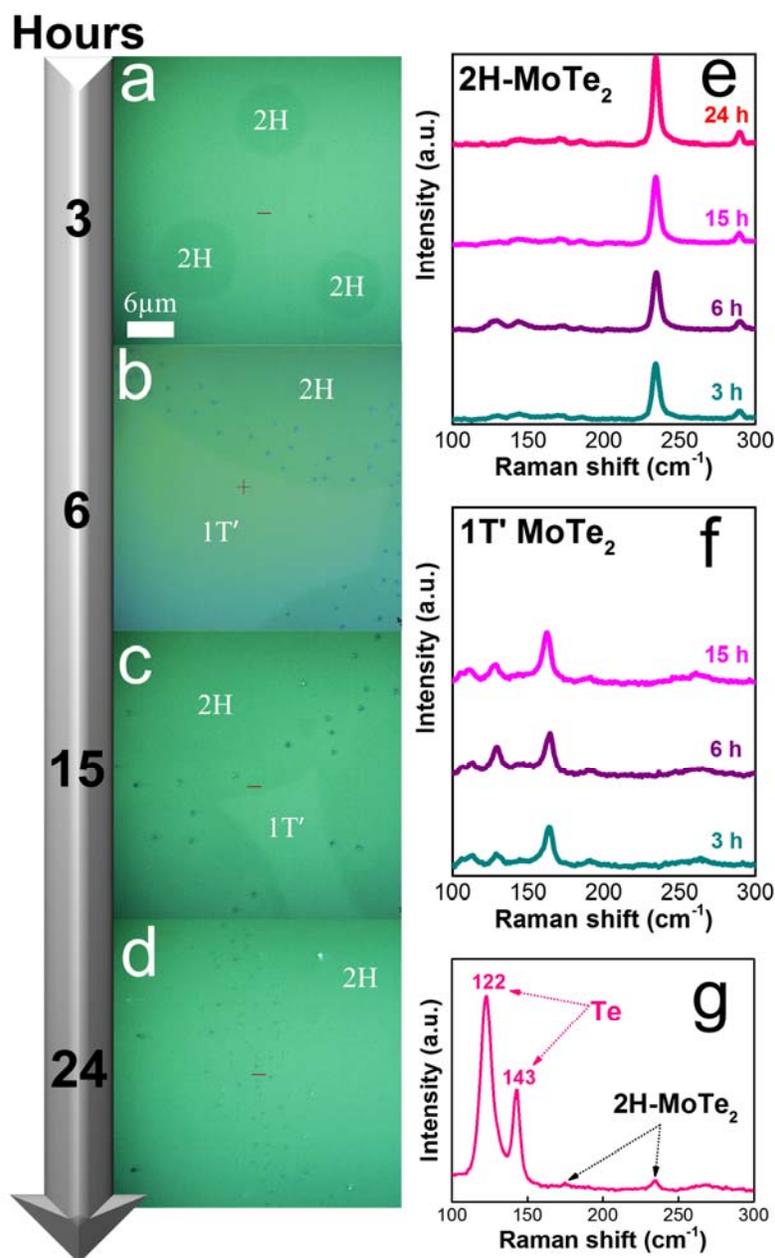

**Figure 2** | Time evolution of the MoTe$_2$ phase during the post-annealing process. (a–d) Top-view optical microscope images of MoTe$_2$ annealed for 3 to 24 h, where 1T′- and 2H-MoTe$_2$ regions displayed an apparent color contrast. The 2H-MoTe$_2$ regions expanded outward from isolated circular islands (a) with the annealing time and merged to form a complete 2H-MoTe$_2$ layer (d). Metallic gray precipitates (b–d) were visible at the 1T′/2H-MoTe$_2$ phase boundaries. (e, f) Typical Raman spectra measured from the 2H- and 1T′ regions, respectively. No 1T′-MoTe$_2$ phase was observed after 24-h annealing. (g) Typical Raman spectrum measured at the regions with metallic gray precipitates, showing characteristic peaks corresponding to Te.

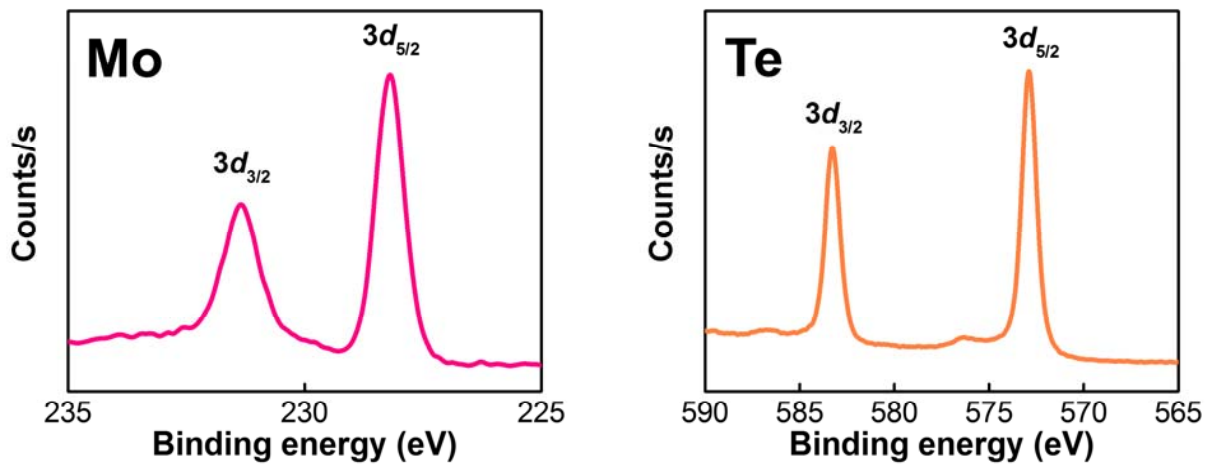

**Figure 3** | XPS analysis of 2H-MoTe$_2$ films annealed at 650 °C for 24 h. The sample exhibited characteristic Mo 3*d* and Te 3*d* binding energies for MoTe$_2$. The sample was decapped using HF etching before the measurement. The Mo/Te composition was 1/2.3. The Te-rich composition could be attributed to the localized Te precipitates.

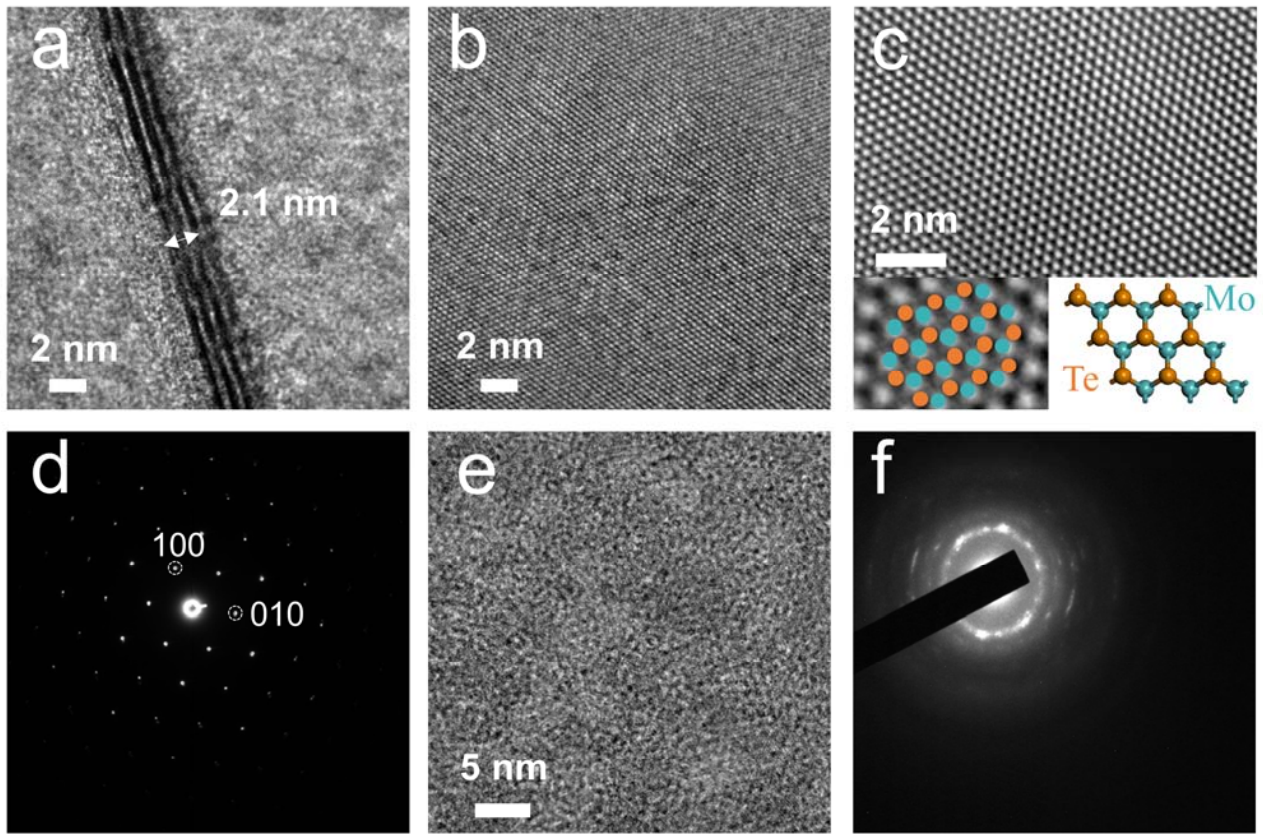

**Figure 4** | Atomic structures of 2H- and 1T′-MoTe$_2$. The 2H- and 1T′-MoTe$_2$ were post-annealed for 24 h and 3 h, respectively. (a) Cross-sectional HRTEM images of 2H-MoTe$_2$, displaying a three-layer structure with a total thickness of 2.1 nm. (b) Plan-view HRTEM of 2H-MoTe$_2$ and (c) its fast-Fourier-transform filtered image, displaying a perfect hexagonal lattice. Inset: corresponding Mo and Te atom arrangement. (d) SAED pattern of single-crystal 2H-MoTe$_2$ with an SAED beam diameter of 650 nm. (e) Plan-view HRTEM and (f) SAED pattern (beam diameter of 260 nm) of polycrystalline 1T′-MoTe$_2$ films.

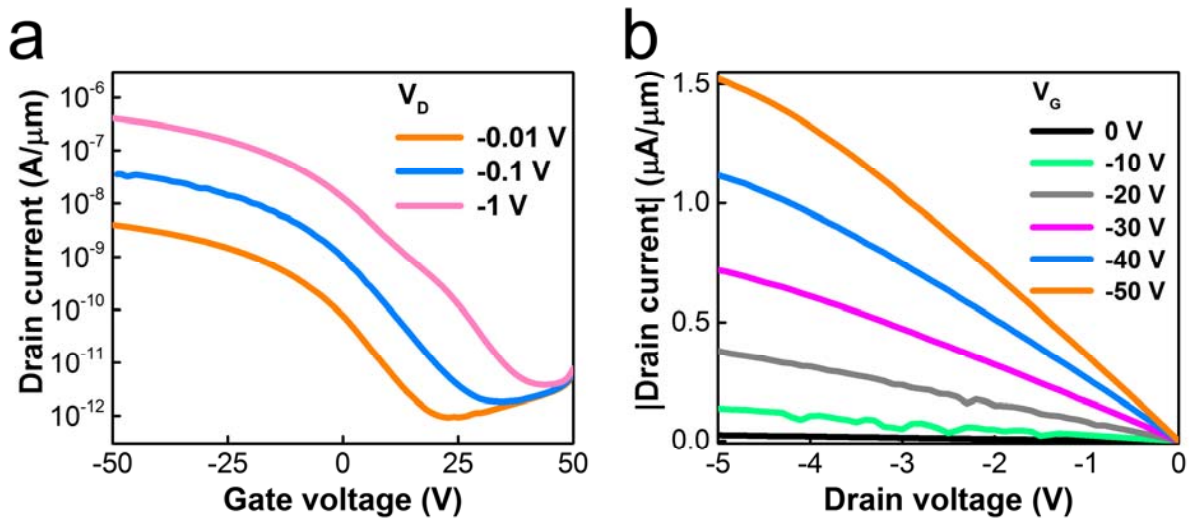

**Figure 5** | Electrical properties of back-gated 2H-MoTe$_2$ FETs with Pd source/drain contacts. (a) Strong $V_D$-dependent transfer characteristics with a current on/off-ratio of $10^5$ at $V_D = -1$ V. (b) Linear output characteristics at small $V_D$ bias for $V_G$ ranging from 0 to $-50$ V. The thickness of the SiO$_2$ gate dielectric was 300 nm, and the device channel was 8 μm long. The drain current was normalized to the channel width.

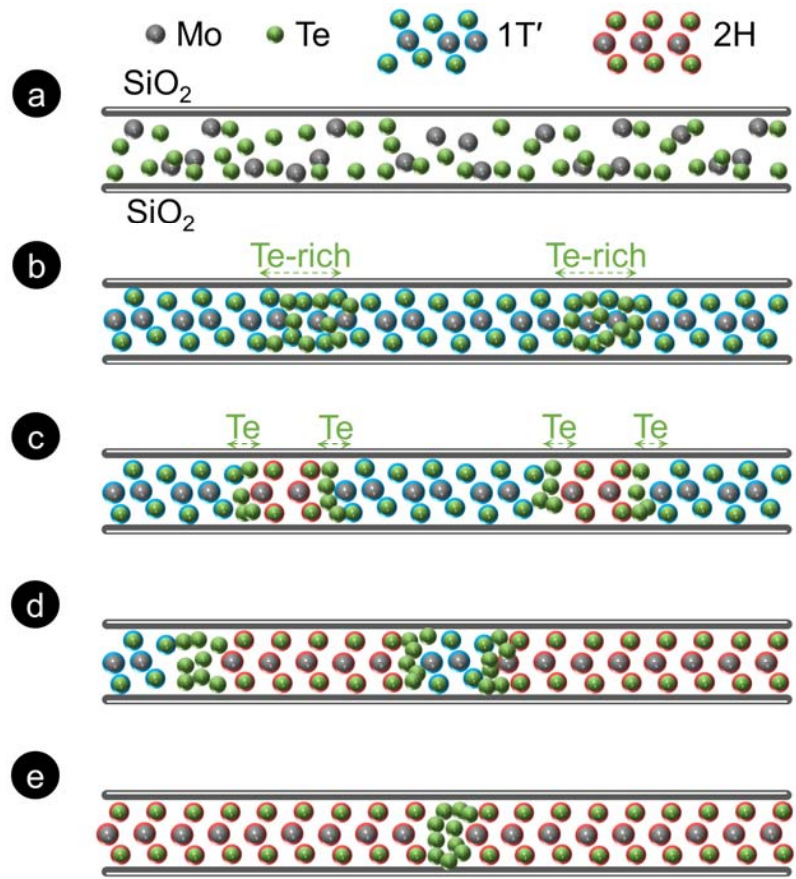

**Figure 6** | Schematics illustrating the time evolution of the recrystallization and phase transition of 2D MoTe$_2$ deposited using PVD and subsequent post-annealing. (a) Low-crystallinity Mo–Te composite deposited using sputtering is encapsulated by the substrate and the SiO$_2$ capping layer. (b) After a short post-annealing period, small-grain polycrystalline 1T′ MoS$_2$ begins to form, and the excessive Te segregates into the amorphous regions between the 1T′ grains. (c) 2H-MoTe$_2$ starts to nucleate in the Te-rich regions. (d) When the annealing time is increased, the regions of recrystallized 2H-MoTe$_2$ expand outward from the nucleation sites as the 1T′-to-2H phase transition is assisted by the excessive-Te forefront. (e) The entire layer is recrystallized into the 2H phase through the merging of 2H-MoTe$_2$ grains, and the excessive Te precipitates at the grain boundaries.

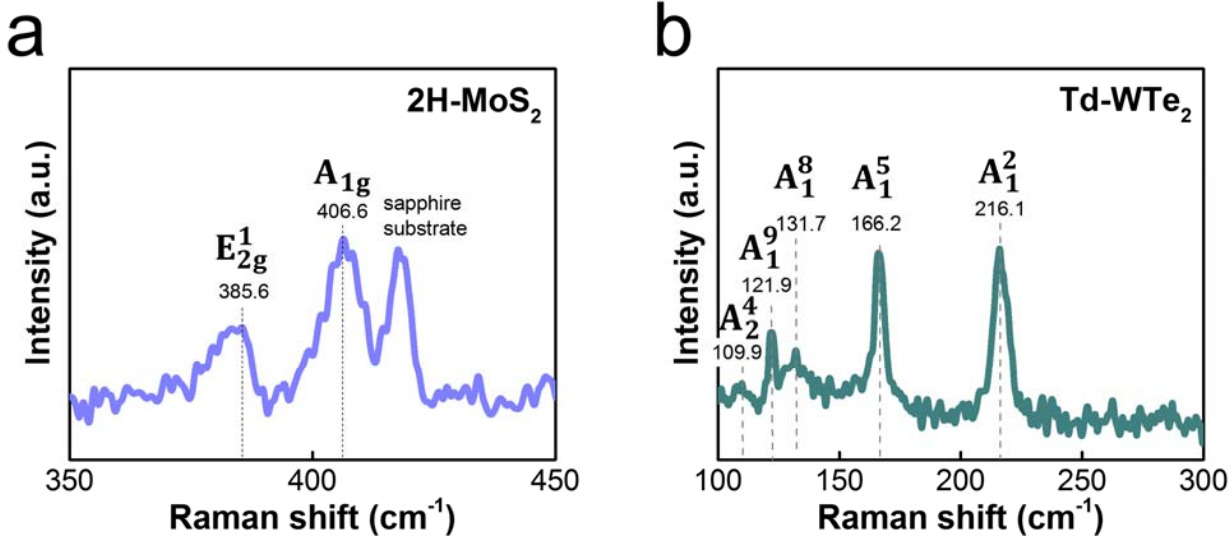

**Figure 7** | 2H-MoS$_2$ and Td-WTe$_2$ deposited using the direct PVD method with SiO$_2$ capping and chalcogen-free post-annealing. Raman spectra confirm (a) semiconductor 2H-MoS$_2$ with characteristic $E_{2g}^1$ and $A_{1g}$ peaks and (b) semimetal Td WTe$_2$ with characteristic $A_1^2$, $A_1^5$, $A_1^8$, $A_1^9$, and $A_2^4$ peaks.

# *Supporting Information*

## Large-Area Two-dimensional Layered MoTe$_2$ by Physical Vapor Deposition and Solid-Phase Crystallization in a Tellurium-Free Atmosphere

Jyun-Hong Huang[1], Kuang-Ying Deng[2], Pang-Shiuan Liu[1], Chien-Ting Wu[3], Cheng-Tung Chou[2], Wen-Hao Chang[4,5], Yao-Jen Lee[3,6], and Tuo-Hung Hou[1,5 a]

[1] Department of Electronics Engineering and Institute of Electronics, National Chiao Tung University, Hsinchu 300, Taiwan
[2] Department of Chemical and Materials Engineering, National Central University, Jhongli 320, Taiwan
[3] National Nano Device Laboratories, Hsinchu 300, Taiwan
[4] Department of Electrophysics, National Chiao Tung University, Hsinchu 300, Taiwan
[5] Taiwan Consortium of Emergent Crystalline Materials, (TCECM), Ministry of Science and Technology, Taipei 106, Taiwan
[6] Department of Physics, National Chung Hsing University, Taichung 402, Taiwan

Correspondence and requests for materials should be addressed to T-H Hou
E-mail: thhou@mail.nctu.edu.tw

## S1. Mechanical exfoliation of MoTe₂ target

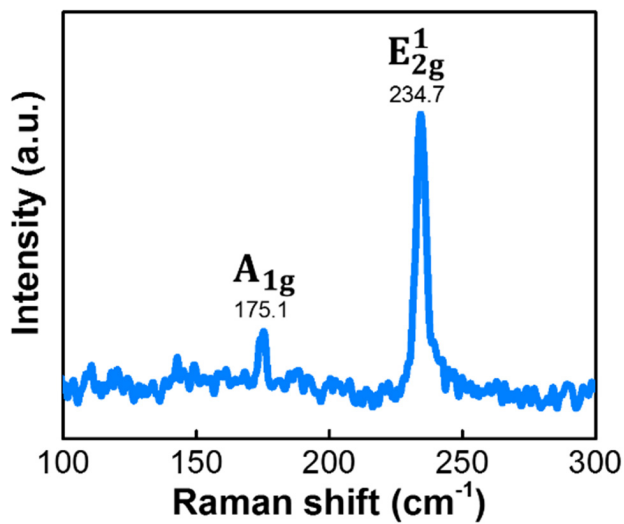

**Figure S1.** Raman spectrum of exfoliated flakes from the MoTe₂ sputter target.

A scotch-tape method was used to exfoliate MoTe₂ flakes from the MoTe₂ target (99.9% purity; Kojundo Chemical Laboratory). The Raman spectrum (Figure S1) exhibited characteristic peaks at 171.4 ($A_{1g}$) and 234.7 ($E_{2g}^1$) cm$^{-1}$, confirming that the film was 2H-MoTe₂. The smaller difference between $A_{1g}$ and $E_{2g}^1$ compared with that observed in Figure 1b and the absence of the $B_{2g}^1$ peak corresponded to bulk 2H-MoTe₂ [S1], indicating that the unoptimized exfoliation method produced thick MoTe₂ flakes.

## S2. Raman spectrum of as-deposited MoTe₂ thin film

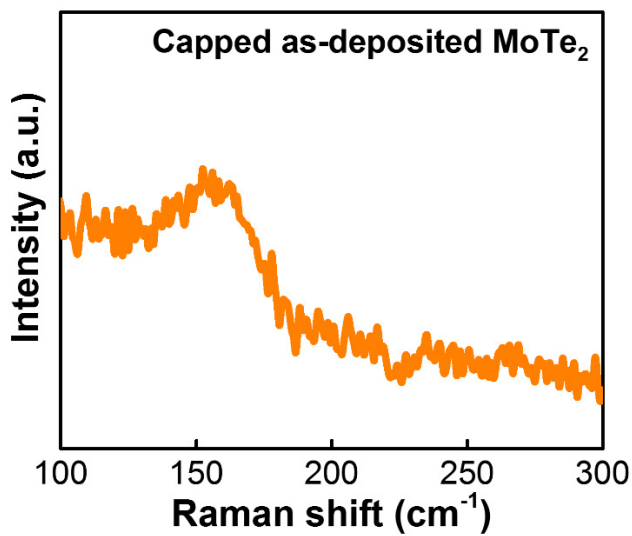

**Figure S2.** Raman spectrum of the capped as-deposited MoTe₂ film.

The as-deposited MoTe₂ film had low crystallinity because the physical ion bombardment involved in sputtering easily broke the weak Mo–Te bonds of the 2H-MoTe₂ target. To prevent possible oxidation when exposed to air, the as-deposited film was immediately capped with an SiO₂ capping layer. The Raman spectrum of the capped as-deposited film had no distinct Raman peaks (Figure S2).

**S3. Effect of MoTe₂ thickness on Raman spectrum**

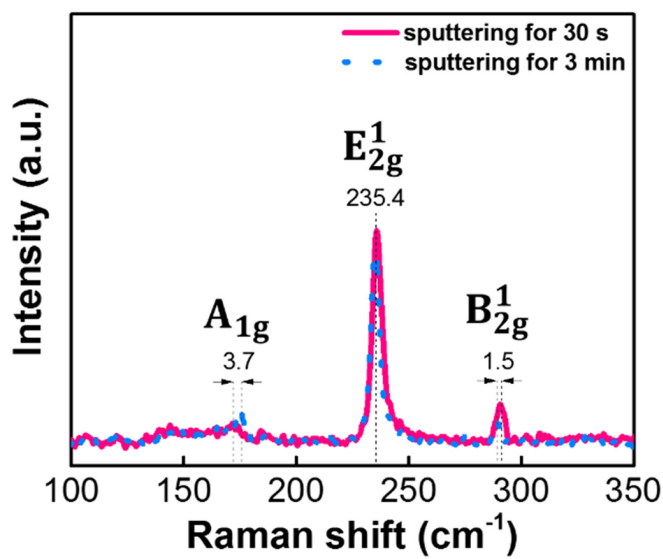

**Figure S3.** Thickness-dependent Raman spectra of 2H-MoTe₂ deposited by the PVD method.

The thickness of the 2H-MoTe₂ film deposited using PVD was controlled by varying the sputtering duration. Although the relationship between number of MoTe₂ layers and the sputtering time should be carefully calibrated in the future, the Raman spectra in Figure S3 illustrate the difference between samples produced with sputtering times of 30 s and 3 min. The spectrum of the thicker sample (3 min) has smaller $A_{1g}$–$E_{2g}^1$ and $B_{2g}^1$–$E_{2g}^1$ differences compared with that of the thinner sample (30 s). The results are consistent with those of previous studies on the thickness-dependent Raman spectra of 2H-MoTe₂ [S1].

## S4. Raman spectrum of the bulk Te flakes

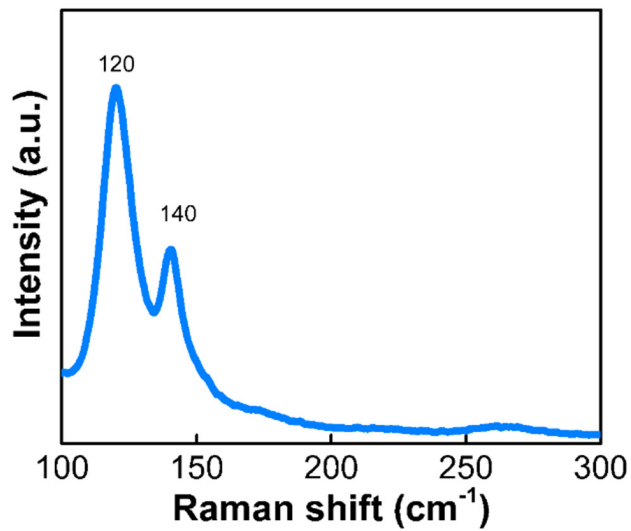

**Figure S4.** Raman spectra of high-purity Te flakes.

To verify the species of metallic grey precipitates generated during the post-annealing process, we measured the Raman spectrum of bulk Te flakes (SIGMA-ALDRICH Co., purity 99.999%) (Figure S4). The characteristic Raman peaks at 120 and 140 cm$^{-1}$ were assigned to the $A_1$ and $E_{TO}$ modes of Te [S2], and were close to those obtained from the precipitates in Figure 2g.

## S5. SiO$_2$ decapping and MoTe$_2$ stability

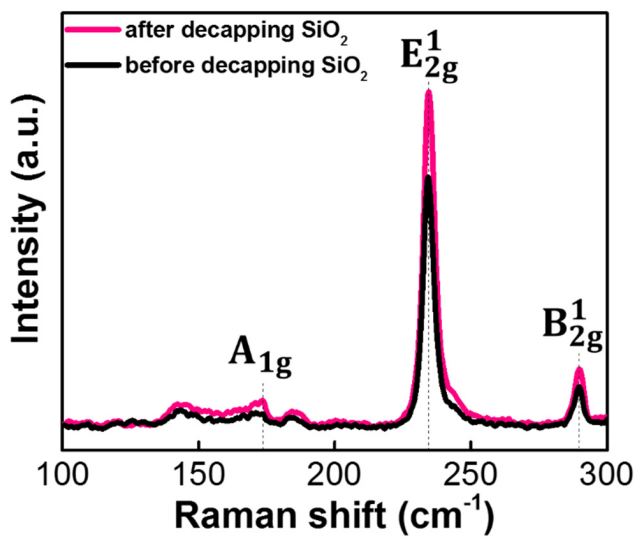

**Figure S5.** Raman spectra of 2H-MoTe$_2$ before and after the SiO$_2$ decapping process.

The 100-nm-thick composite SiO$_2$ capping layer was removed by dipping the sample in diluted hydrogen fluoride (HF) (0.58%) for 8 min. Figure S5 presents a comparison of the Raman spectra before and after the SiO$_2$ decapping. Nearly intact Raman spectra indicate that the decapping process did not degrade the quality of 2H-MoTe$_2$.

## S6. Long-term stability of the decapped 2H-MoTe$_2$

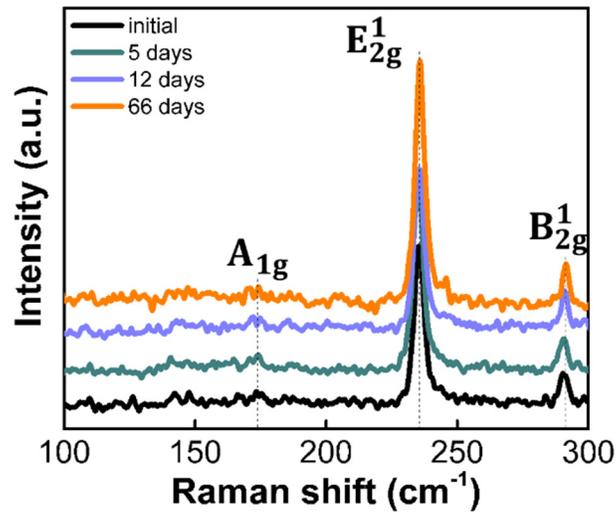

**Figure S6.** Time evaluation of Raman spectra of the decapped 2H-MoTe$_2$.

The decapped 2H-MoTe$_2$ was stored in a vacuum desiccator containing low moisture levels. The sample remained extremely stable for more than 2 months, as the intact Raman spectra in Figure S6 demonstrates.

## S7. Effect of Mo and Pd contacts on 2H-MoTe$_2$ back-gated FETs

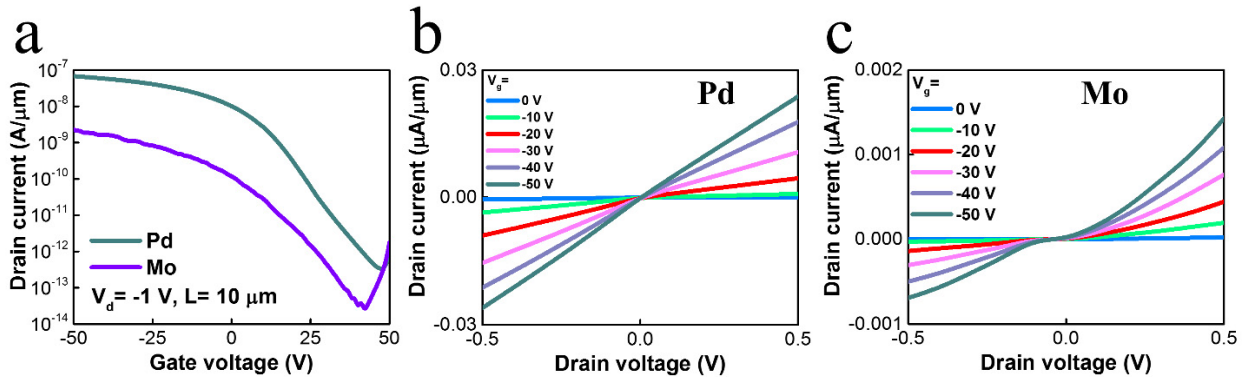

**Figure S7.** Comparison of (a) transfer curves and (b), (c) output curves of 2H-MoTe$_2$ back-gated FETs with Pd and Mo contacts.

Metal source/drain contacts often form Schottky contacts with 2D TMDs and substantially affect current transport in FET devices. Figure S7 presents a comparison of two metal contact materials, Mo and Pd, in the 2H-MoTe$_2$ back-gated FET devices. The sample with Pd contacts exhibited superior p-type conduction characteristics and linear output conductance at small $V_D$, possibly because of its higher work function (5.2 eV for Pd vs. 4.6 eV for Mo).

***References:***

[S1] M. Yamamoto, S. T. Wang, M. Ni, Y. F. Lin, S. L. Li, S. Aikawa, W. B. Jian, K. Ueno, K. Wakabayashi, K. Tsukagoshi, *ACS Nano* **2014**, 8, 3895.

[S2] A. S. Pine, G. Dresselhaus, *Phys. Rev. B* **1971**, 4, 356.